\documentclass[prl,twocolumn,nofootinbib,showpacs]{revtex4}

\begin{document}

\title{Geometrical formulation of classical electromagnetism} 
\author{Nikodem J. Pop\l awski}
\affiliation{Department of Physics, Indiana University,
727 East Third Street, Bloomington, Indiana 47405, USA}
\date{\today}

\begin{abstract}
A general affine connection has enough degrees of freedom to describe the classical gravitational and electromagnetic fields in the metric-affine formulation of gravity.
The gravitational field is represented in the Lagrangian by the symmetric part of the Ricci tensor, while the classical electromagnetic field is represented geometrically by the tensor of homothetic curvature.
We introduce matter as the four-velocity field subject to the kinematical constraint in which the Lagrange multiplier represents the energy density.
A coupling between the four-velocity and the trace of the nonmetricity tensor represents the electric charge density.
We show that the simplest metric-affine Lagrangian that depends on the Ricci tensor and the tensor of homothetic curvature generates the Einstein-Maxwell field equations, while the Bianchi identity gives the Lorentz equation of motion.
If the four-velocity couples to the torsion vector, the Einstein equations are modified by a term that is significant at the Planck scale and may prevent the formation of spacetime singularities.
\end{abstract}

\maketitle

{\em Introduction}.
The geometry of general relativity is given by a 4D Riemannian manifold, that is, with a symmetric metric tensor and an affine connection that is torsionless and metric compatible.
The electromagnetic field and its sources are considered to be on the side of the matter tensor in the field equations, acting as sources of the gravitational field.
In unified field theories, the electromagnetic field obtains the geometrical status as the gravitational field~\cite{Goe} by modifying some postulates of general relativity.
Weyl relaxed the metric compatibility of the affine connection, obtaining a theory, where electromagnetic gauge transformation was related to conformal transformation of the metric~\cite{Weyl}.
Kaluza showed that the Lagrangian proportional to the Ricci scalar in a 5D spacetime yields the Einstein-Maxwell field equations and the Lorentz equation of motion~\cite{Kaluza}.
Relaxing the symmetry of the connection and metric tensor resulted in the Einstein-Straus and Schr\"{o}dinger nonsymmetric field theories~\cite{Schrod,Schr,Einst}.
While Kaluza's theory gave rise to later models with extra dimensions, the Einstein-Schr\"{o}dinger and Weyl theories turned out to be unphysical~\cite{Lord}.

In the metric-affine formulation of gravity, both the metric and connection are independent variables, and the field equations are derived by varying the action with respect to these quantities~\cite{Einst,MAE,MA1,MA2}.
Ponomarev and Obukhov showed that a linear connection, not restricted to be metric compatible and symmetric, has enough degrees of freedom to describe the gravitational and electromagnetic fields~\cite{PO}.
While the gravitational field is represented by the symmetric part of the Ricci tensor, the classical electromagnetic field can be represented by the tensor of homothetic curvature~\cite{Scho}.
The simplest metric-affine Lagrangian that depends on both tensors is linear in the Ricci scalar and quadratic in the tensor of homothetic curvature, generating the Einstein-Maxwell equations with sources~\cite{PO,NikoA} represented by the hypermomentum density~\cite{MA2}.
The analogous purely affine Lagrangian was introduced by Ferraris and Kijowski~\cite{FK}.
The invariance of the action under Einstein's $\lambda$-transformations of the connection~\cite{Einst} corresponds to gauge invariance in electromagnetism~\cite{PO}.
Unlike in~\cite{Schrod,Schr,Einst}, where the electromagnetic field is associated with the metric, interpreting electromagnetism as a part of the affine connection seems more natural.
While the connection generalizes an ordinary derivative of a vector into a coordinate-covariant derivative, the electromagnetic potential generalizes it into a $U(1)$-covariant derivative.

The previous unified field theories, except Kaluza's theory, did not derive the Lorentz equation of motion for charged particles in purely geometrical way from a variational principle.
In this paper we show that such a derivation is possible if we use the Lagrange-multiplier method of Taub~\cite{Taub}, introducing continuous matter in the Lagrangian as the four-velocity field subject to a kinematical constraint.
The Lagrange multiplier in this constraint turns out to represent the energy density.
Unlike in~\cite{Taub}, we do not regard the four-velocity as a dynamical variable; we only vary the metric, the connection, and the Lagrange multiplier.
We represent the electromagnetic field by the tensor of homothetic curvature~\cite{PO}, and the electric charge density as a coupling between the four-velocity and the trace of the nonmetricity tensor.
We show that this four-term Lagrangian generates the Einstein equations from varying the metric, and the Maxwell equations from varying the connection.
The Bianchi identity applied to the Einstein equations yields the Lorentz equation of motion.
If the four-velocity couples to the torsion vector, the Einstein equations are modified by a term that is significant at the Planck scale and may prevent the formation of singularities.

{\em Lagrangian}.
A general metric-affine Lagrangian density ${\cal L}$ depends on the affine connection $\Gamma^{\,\,\rho}_{\mu\,\nu}$ and the curvature tensor, $R^\rho_{\phantom{\rho}\mu\sigma\nu}=\Gamma^{\,\,\rho}_{\mu\,\nu,\sigma}-\Gamma^{\,\,\rho}_{\mu\,\sigma,\nu}+\Gamma^{\,\,\kappa}_{\mu\,\nu}\Gamma^{\,\,\rho}_{\kappa\,\sigma}-\Gamma^{\,\,\kappa}_{\mu\,\sigma}\Gamma^{\,\,\rho}_{\kappa\,\nu}$, as well as the symmetric metric tensor $g_{\mu\nu}$ of the Lorentzian signature $(+,-,-,-)$.
The antisymmetric part of the affine connection is the Cartan torsion tensor, $S^\rho_{\phantom{\rho}\mu\nu}=\Gamma^{\,\,\,\,\rho}_{[\mu\,\nu]}$, while its symmetric part can be split into the Christoffel symbols, $\{^{\,\,\rho}_{\mu\,\nu}\}=\frac{1}{2}g^{\rho\lambda}(g_{\nu\lambda,\mu}+g_{\mu\lambda,\nu}-g_{\mu\nu,\lambda})$, and terms constructed from the torsion tensor and the nonmetricity tensor, $N_{\mu\nu\rho}=g_{\mu\nu;\rho}$~\cite{MA1,MA2}: $\Gamma^{\,\,\rho}_{\mu\,\nu}=\{^{\,\,\rho}_{\mu\,\nu}\}+S^\rho_{\phantom{\rho}\mu\nu}+2S_{(\mu\nu)}^{\phantom{(\mu\nu)}\rho}+\frac{1}{2}N_{\mu\nu}^{\phantom{\mu\nu}\rho}-N^\rho_{\phantom{\rho}(\mu\nu)}$, where the semicolon denotes the covariant differentiation with respect to $\Gamma^{\,\,\rho}_{\mu\,\nu}$~\cite{Schr,LL}.
We assume that the dependence of ${\cal L}$ on the curvature is restricted to the contracted curvature tensors: the Ricci tensor, $R_{\mu\nu}=R^\rho_{\phantom{\rho}\mu\rho\nu}$, and the antisymmetric tensor of homothetic curvature:
\begin{equation}
Q_{\mu\nu}=R^\rho_{\phantom{\rho}\rho\mu\nu}=\Gamma^{\,\,\rho}_{\rho\,\nu,\mu}-\Gamma^{\,\,\rho}_{\rho\,\mu,\nu},
\label{homoth}
\end{equation}
which has the form of a curl of a vector~\cite{Schr,Scho}: $Q_{\mu\nu}=-\frac{1}{2}(N^\rho_{\phantom{\rho}\rho\nu,\mu}-N^\rho_{\phantom{\rho}\rho\mu,\nu})$.

The simplest metric-affine Lagrangian density that depends on $R_{\mu\nu}$ is the Einstein-Hilbert Lagrangian density for the gravitational field~\cite{MAE}:
\begin{equation}
{\cal L}_g=-\frac{1}{2\kappa}R_{\mu\nu}{\sf g}^{\mu\nu},
\label{LagGrav}
\end{equation}
where $\kappa=8\pi G$ ($c=1$), ${\sf g}^{\mu\nu}=\sqrt{-g}g^{\mu\nu}$ is the fundamental metric density~\cite{Schr}, and $g=\mbox{det}(g_{\mu\nu})$.
The total Lagrangian density for the gravitational field and of matter is given by ${\cal L}_g+{\cal L}_Q+{\cal L}_m$, where the Lagrangian density for matter ${\cal L}_m$ depends in general on both the metric and connection.
The variations of ${\cal L}_m$ with respect to the metric and connection define, respectively, the dynamical energy-momentum tensor $T_{\mu\nu}=\frac{2}{\sqrt{-g}}\frac{\delta{\cal L}_m}{\delta g^{\mu\nu}}$, and the hypermomentum density~\cite{MA2}:
\begin{equation}
\Pi_{\phantom{\mu}\rho\phantom{\nu}}^{\mu\phantom{\rho}\nu}=-2\kappa\frac{\delta{\cal L}_m}{\delta \Gamma^{\,\,\rho}_{\mu\,\nu}},
\label{Pi}
\end{equation}
which has the same dimension as the connection.

The metric-affine theory based on the Lagrangian density ${\cal L}_g$ does not determine the connection uniquely because ${\cal L}_g$ is invariant under projective transformations~\cite{MA2,PO}:
\begin{equation}
\Gamma^{\,\,\rho}_{\mu\,\nu}\rightarrow\Gamma^{\,\,\rho}_{\mu\,\nu}+\delta^\rho_\mu V_\nu,
\label{proj}
\end{equation}
where $V_\nu$ is a vector function of the coordinates.
The same problem occurs if we add to ${\cal L}_g$ a Lagrangian for matter ${\cal L}_m$ that does not depend on the connection, for example, representing the electromagnetic field or an ideal fluid.
Therefore at least four degrees of freedom must be constrained to make such a theory consistent from a physical point of view~\cite{MA2}.
If, however, ${\cal L}_m$ does depend on the connection, for example, for spinor fields, the projective invariance of ${\cal L}_g+{\cal L}_m$ imposes four algebraic constraints on $\Pi_{\phantom{\mu}\rho\phantom{\nu}}^{\mu\phantom{\rho}\nu}$ and restricts forms of matter that can be described by metric-affine gravity~\cite{var}.
Consequently, we must add to ${\cal L}_g+{\cal L}_m$ a term that is not projectively invariant, so the total Lagrangian density will be projectively invariant.

The Lagrangian density that depends on the tensor of homothetic curvature is an example of such a term that is purely geometrical.
The tensor $Q_{\mu\nu}$ is invariant with respect to projective transformations with $V_\nu=\lambda_{,\nu}$, where $\lambda$ is a scalar function of the coordinates ($\lambda$-transformations)~\cite{Einst}.
The simplest metric-affine Lagrangian density that depends on $Q_{\mu\nu}$, proposed by Ponomarev and Obukhov~\cite{PO}, has the form of the Maxwell Lagrangian for the electromagnetic field:
\begin{equation}
{\cal L}_Q=\frac{\alpha^2}{4}\sqrt{-g}Q_{\mu\nu}Q^{\mu\nu},
\label{LagEM}
\end{equation}
where $\alpha$ is a constant.

{\em Field equations}.
We consider the total Lagrangian density ${\cal L}={\cal L}_g+{\cal L}_Q+{\cal L}_m$.
From the stationarity of the action $S=\int d^4x{\cal L}$ under arbitrary variations of $g^{\mu\nu}$: $\delta S=0$, we obtain the metric-affine Einstein equations:
\begin{equation}
R_{(\mu\nu)}-\frac{1}{2}Rg_{\mu\nu}=\kappa T_{\mu\nu}-\kappa\alpha^2\Bigl(\frac{1}{4}Q_{\rho\sigma}Q^{\rho\sigma}g_{\mu\nu}-Q_{\mu\rho}Q_\nu^{\phantom{\nu}\rho}\Bigr),
\label{Ein1}
\end{equation}
where $R=R_{\mu\nu}g^{\mu\nu}$ is the Ricci scalar.

The variation of $S$ with respect to the affine connection $\Gamma^{\,\,\rho}_{\mu\,\nu}$ is $\delta S=-\frac{1}{2\kappa}\int d^4x({\sf g}^{\mu\nu}\delta R_{\mu\nu}+{\sf h}^{\mu\nu}\delta Q_{\mu\nu}+\Pi_{\phantom{\mu}\rho\phantom{\nu}}^{\mu\phantom{\rho}\nu}\delta\Gamma^{\,\,\rho}_{\mu\,\nu})$, where the antisymmetric tensor density ${\sf h}^{\mu\nu}=-2\kappa\frac{\delta{\cal L}_Q}{\delta Q_{\mu\nu}}$ is linear in the tensor of homothetic curvature~\cite{NikoA,NikoI}:
\begin{equation}
{\sf h}^{\mu\nu}=-\kappa\alpha^2\sqrt{-g}Q^{\mu\nu}.
\label{as}
\end{equation}
The variation of the Ricci tensor is given by the Palatini formula~\cite{Schr,Scho}: $\delta R_{\mu\nu}=\delta\Gamma^{\,\,\rho}_{\mu\,\nu;\rho}-\delta\Gamma^{\,\,\rho}_{\mu\,\rho;\nu}-2S^\sigma_{\phantom{\sigma}\rho\nu}\delta\Gamma^{\,\,\rho}_{\mu\,\sigma}$.
Using the identity $\int d^4x({\sf V}^\mu)_{;\mu}=2\int d^4x S_\mu{\sf V}^\mu$, where ${\sf V}^\mu$ is an arbitrary vector density and $S_\mu=S^\nu_{\phantom{\nu}\mu\nu}$ is the torsion vector, from the stationarity of the action under arbitrary variations of $\Gamma^{\,\,\rho}_{\mu\,\nu}$ we obtain
\begin{eqnarray}
& & {\sf g}^{\mu\nu}_{\phantom{\mu\nu};\rho}-{\sf g}^{\mu\sigma}_{\phantom{\mu\sigma};\sigma}\delta^\nu_\rho-2{\sf g}^{\mu\nu}S_\rho+2{\sf g}^{\mu\sigma}S_\sigma\delta^\nu_\rho+2{\sf g}^{\mu\sigma}S^\nu_{\phantom{\nu}\rho\sigma} \nonumber \\
& & =\Pi_{\phantom{\mu}\rho\phantom{\nu}}^{\mu\phantom{\rho}\nu}+2{\sf h}^{\nu\sigma}_{\phantom{\nu\sigma},\sigma}\delta^\mu_\rho.
\label{field1}
\end{eqnarray}

Contracting the indices $\mu$ and $\rho$ in Eq.~(\ref{field1}) gives the Maxwell-Minkowski-like equation:
\begin{equation}
{\sf h}^{\sigma\nu}_{\phantom{\sigma\nu},\sigma}={\sf j}^\nu,
\label{Max1}
\end{equation}
with the source represented by the trace of the hypermomentum density:
\begin{equation}
{\sf j}^\nu=\frac{1}{8}\Pi_{\phantom{\sigma}\sigma\phantom{\nu}}^{\sigma\phantom{\sigma}\nu}.
\label{Max2}
\end{equation}
Since the tensor density ${\sf h}^{\mu\nu}$ is antisymmetric, the current vector density~(\ref{Max2}) is conserved: ${\sf j}^\mu_{\phantom{\mu},\mu}=0$, which constrains how the connection $\Gamma^{\,\,\rho}_{\mu\,\nu}$ can enter the metric-affine Lagrangian density for matter ${\cal L}_m$: $\Pi_{\phantom{\sigma}\sigma\phantom{\nu},\nu}^{\sigma\phantom{\sigma}\nu}=0$.

If the field-part of the Lagrangian does not depend on $Q_{\mu\nu}$ (or other geometrical quantities beside the affine connection and the symmetrized Ricci tensor), the field equation~(\ref{Max1}) becomes a stronger, algebraic constraint on how the matter-part of the Lagrangian depends on the connection: $\Pi_{\phantom{\sigma}\sigma\phantom{\nu}}^{\sigma\phantom{\sigma}\nu}=0$, which restricts forms of matter that can be described by the metric-affine formulation of gravity~\cite{MA2,var}.
The dependence of a metric-affine Lagrangian on the tensor of homothetic curvature $Q_{\mu\nu}$ replaces this unphysical constraint with the field equation for ${\sf h}^{\mu\nu}$.
If this dependence is given by~(\ref{LagEM}), the field equation from varying the affine connection has the form of the Maxwell equations.

The tensor $R_{(\mu\nu)}$ is invariant under an infinitesimal projective transformation~(\ref{proj}): $\delta\Gamma^{\,\,\rho}_{\mu\,\nu}=\delta^\rho_\mu\delta V_\nu$.
Under the same transformation, the tensor $Q_{\mu\nu}$ changes according to $Q_{\mu\nu}\rightarrow Q_{\mu\nu}+4(\delta V_{\nu,\mu}-\delta V_{\mu,\nu})$.
Consequently, the action changes according to $\delta S=-\frac{1}{2\kappa}\int d^4x(\Pi_{\phantom{\mu}\rho\phantom{\nu}}^{\mu\phantom{\rho}\nu}\delta\Gamma^{\,\,\rho}_{\mu\,\nu}+{\sf h}^{\mu\nu}\delta Q_{\mu\nu})=-\frac{1}{2\kappa}\int d^4x(\Pi_{\phantom{\sigma}\sigma\phantom{\mu}}^{\sigma\phantom{\sigma}\mu}+8{\sf h}^{\mu\nu}_{\phantom{\mu\nu},\nu})\delta V_\mu$.
This expression is identically zero due to the field equation~(\ref{Max1}), so, although ${\cal L}_m$ and ${\cal L}_Q$ are not projectively invariant, the total action is.

{\em Electromagnetism}.
The formal similarity between the tensor of homothetic curvature $Q_{\mu\nu}$ and the electromagnetic field tensor $F_{\mu\nu}=A_{\nu,\mu}-A_{\mu,\nu}$ (both tensors are curls) suggests that they are proportional to one another~\cite{PO,FK}:
\begin{equation}
Q_{\mu\nu}=\frac{i}{\alpha}F_{\mu\nu}.
\label{prop1}
\end{equation}
Accordingly, the trace of the nonmetricity tensor is proportional to the electromagnetic four-potential $A_\mu$:
\begin{equation}
N^\rho_{\phantom{\rho}\rho\mu}=-\frac{2i}{\alpha}A_\mu.
\label{prop2}
\end{equation}
The $\lambda$-transformation of the connection is thus a geometrical representation of the gauge transformation of the electromagnetic potential: $A_\mu \rightarrow A_\mu+\phi_{,\mu}$, with $\phi=-4i\alpha\lambda$~\cite{PO}.
Eqs.~(\ref{as}), (\ref{Max1}) and~(\ref{prop1}), and the Maxwell equations: $F^{\mu\nu}_{\phantom{\mu\nu}:\mu}=j^\nu$, where the colon denotes the covariant differentiation with respect to $\{^{\,\,\rho}_{\mu\,\nu}\}$, relate the electromagnetic current vector $j^\nu$ to the trace of the hypermomentum density ${\sf j}^\nu$:
\begin{equation}
j^\nu=\frac{i}{\kappa\alpha\sqrt{-g}}{\sf j}^\nu.
\label{prop3}
\end{equation}

Associating $Q_{\mu\nu}$ with $F_{\mu\nu}$ allows us to interpret the electromagnetic field as the field whose purpose is to remove the unphysical algebraic constraint on the hypermomentum density in metric-affine gravity.
The reason for the appearance of $i$ in Eqs.~(\ref{prop1}) and~(\ref{prop2}) is to assure the covariant derivative, containing the nonmetricity tensor (which is a part of the connection), coincide with the covariant derivative of the $U(1)$ gauge symmetry, containing the electromagnetic term $ieA_\mu$ ($\hbar=1$).
Eq.~(\ref{prop2}) indicates that $\alpha$ is inverse proportional to the charge of the electron.
We can show, using the coupling of the affine connection to spinors in the tetrad formalism~\cite{st}, that $\alpha=\frac{1}{4e}$~\cite{NikoG}.

{\em Solution of field equations}.
Eq.~(\ref{field1}) is equivalent to
\begin{eqnarray}
& & {\sf g}^{\mu\nu}_{\phantom{\mu\nu},\rho}+\,^\ast\Gamma^{\,\,\mu}_{\sigma\,\rho}{\sf g}^{\sigma\nu}+\,^\ast\Gamma^{\,\,\nu}_{\rho\,\sigma}{\sf g}^{\mu\sigma}-\,^\ast\Gamma^{\,\,\sigma}_{\sigma\,\rho}{\sf g}^{\mu\nu} \nonumber \\
& & =\Pi_{\phantom{\mu}\rho\phantom{\nu}}^{\mu\phantom{\rho}\nu}-\frac{1}{3}\Pi_{\phantom{\mu}\sigma\phantom{\sigma}}^{\mu\phantom{\sigma}\sigma}\delta^\nu_\rho+2{\sf h}^{\nu\sigma}_{\phantom{\nu\sigma},\sigma}\delta^\mu_\rho-\frac{2}{3}{\sf h}^{\mu\sigma}_{\phantom{\mu\sigma},\sigma}\delta^\nu_\rho,
\label{field2}
\end{eqnarray}
where $^\ast\Gamma^{\,\,\rho}_{\mu\,\nu}=\Gamma^{\,\,\rho}_{\mu\,\nu}+\frac{2}{3}\delta^\rho_\mu S_\nu$ is the projectively invariant part of the affine connection (Schr\"{o}dinger's star-affinity)~\cite{Schr}.

Substituting Eqs.~(\ref{Max1}) and~(\ref{Max2}) to~(\ref{field2}) gives a linear relation between $^\ast\Gamma^{\,\,\rho}_{\mu\,\nu}$ and the hypermomentum density $\Pi_{\phantom{\mu}\rho\phantom{\nu}}^{\mu\phantom{\rho}\nu}$.
If we decompose the star-affinity $^\ast\Gamma^{\,\,\rho}_{\mu\,\nu}$ as~\cite{NikoA}:
\begin{equation}
^\ast\Gamma^{\,\,\rho}_{\mu\,\nu}=\{^{\,\,\rho}_{\mu\,\nu}\}+V^\rho_{\phantom{\rho}\mu\nu},
\label{dec}
\end{equation}
where $V^\rho_{\phantom{\rho}\mu\nu}$ is a projectively invariant deflection tensor, then the general solution for $V^\rho_{\phantom{\rho}\mu\nu}$ is given by~\cite{NikoA} (cf. also~\cite{PO}):
\begin{eqnarray}
& & V^\rho_{\phantom{\rho}\mu\nu}=\frac{1}{2\sqrt{-g}}(\Delta_{\phantom{\rho}\nu\phantom{\sigma}}^{\rho\phantom{\nu}\sigma}g_{\mu\sigma}+\Delta_{\phantom{\rho}\mu\phantom{\sigma}}^{\rho\phantom{\mu}\sigma}g_{\nu\sigma}-\Delta_{\phantom{\alpha}\gamma\phantom{\beta}}^{\alpha\phantom{\gamma}\beta}g_{\mu\alpha}g_{\nu\beta}g^{\rho\gamma} \nonumber \\
& & +\Omega_\nu^{\phantom{\nu}\rho\sigma}g_{\mu\sigma}-\Omega_\mu^{\phantom{\mu}\rho\sigma}g_{\nu\sigma}-\Omega_\gamma^{\phantom{\gamma}\alpha\beta}g_{\mu\alpha}g_{\nu\beta}g^{\rho\gamma}),
\label{sol}
\end{eqnarray}
where: $\Delta_{\phantom{\mu}\rho\phantom{\nu}}^{\mu\phantom{\rho}\nu}=\Sigma_{\phantom{\mu}\rho\phantom{\nu}}^{\mu\phantom{\rho}\nu}-\frac{1}{2}\Sigma_{\phantom{\alpha}\rho\phantom{\beta}}^{\alpha\phantom{\rho}\beta}g_{\alpha\beta}g^{\mu\nu}$, $\Sigma_{\phantom{\mu}\rho\phantom{\nu}}^{\mu\phantom{\rho}\nu}=\Pi_{\phantom{(\mu}\rho\phantom{\nu)}}^{(\mu\phantom{\rho}\nu)}-\frac{1}{3}\delta^{(\mu}_\rho\Pi_{\phantom{\nu)}\sigma\phantom{\sigma}}^{\nu)\phantom{\sigma}\sigma}-\frac{1}{6}\Pi_{\phantom{\sigma}\sigma\phantom{(\mu}}^{\sigma\phantom{\sigma}(\mu}\delta^{\nu)}_\rho$, and $\Omega_\rho^{\phantom{\rho}\mu\nu}=\Pi_{\phantom{[\mu}\rho\phantom{\nu]}}^{[\mu\phantom{\rho}\nu]}-\frac{1}{3}\Pi_{\phantom{[\sigma}\sigma\phantom{\nu]}}^{[\sigma\phantom{\sigma}\nu]}\delta^\mu_\rho+\frac{1}{3}\Pi_{\phantom{[\sigma}\sigma\phantom{\mu]}}^{[\sigma\phantom{\sigma}\mu]}\delta^\nu_\rho$.
For the connection given by Eq.~(\ref{dec}), the Ricci tensor is quadratic in $V^\rho_{\phantom{\rho}\mu\nu}$~\cite{Scho}, that is, in $\Pi_{\phantom{\mu}\rho\phantom{\nu}}^{\mu\phantom{\rho}\nu}$: $R_{\mu\nu}=R_{\mu\nu}^{(g)}-\frac{4}{3}S_{[\nu:\mu]}+2V^\rho_{\phantom{\rho}\mu[\nu:\rho]}+V^\sigma_{\phantom{\sigma}\mu\nu}V^\rho_{\phantom{\rho}\sigma\rho}-V^\sigma_{\phantom{\sigma}\mu\rho}V^\rho_{\phantom{\rho}\sigma\nu}$, where $R_{\mu\nu}^{(g)}$ is the Riemannian Ricci tensor constructed from the connection $\Gamma^{\,\,\rho}_{\mu\,\nu}=\{^{\,\,\rho}_{\mu\,\nu}\}$.
Substituting $R_{\mu\nu}$ to the symmetrized Eq.~(\ref{Ein1}) and moving the terms with $V^\rho_{\phantom{\rho}\mu\nu}$ to the right-hand side gives the metric Einstein equations.

The Ricci scalar is:
\begin{equation}
R=R_g+V^{\rho\sigma}_{\phantom{\rho\sigma}\sigma:\rho}-V_{\rho\phantom{\rho\sigma}:\sigma}^{\phantom{\rho}\rho\sigma}+V^{\sigma\lambda}_{\phantom{\sigma\lambda}\lambda}V^\rho_{\phantom{\rho}\sigma\rho}-V_{\sigma\lambda\rho}V^{\rho\sigma\lambda},
\label{Ein2}
\end{equation}
where $R_g=R_{\alpha\beta}^{(g)}g^{\alpha\beta}$ is the Riemannian curvature scalar.
The tensor of homothetic curvature is~\cite{Scho}: $Q_{\mu\nu}=-\frac{8}{3}(S_{\nu,\mu}-S_{\mu,\nu})+V^\rho_{\phantom{\rho}\rho\nu,\mu}-V^\rho_{\phantom{\rho}\rho\mu,\nu}$, which, because of Eq.~(\ref{prop1}), relates the complex torsion vector to the electromagnetic potential and the trace of the deflection tensor:
\begin{equation}
S_\nu=\frac{3}{8}\Bigl(-\frac{iA_\nu}{\alpha}+V^\rho_{\phantom{\rho}\rho\nu}\Bigr).
\label{tor}
\end{equation}
If there are no sources ($\Pi_{\phantom{\mu}\rho\phantom{\nu}}^{\mu\phantom{\rho}\nu}=0$), the connection depends only on $g_{\mu\nu}$ representing the gravitational field, and the torsion vector $S_\mu$ proportional to $A_\mu$~\cite{Ham}.

{\em Matter Lagrangian and equation of motion}.
We regard matter as the four-velocity field $u^\mu=\frac{dx^\mu}{ds}$ that enters the Lagrangian in a kinematical constraint parametrizing the length of a world line $s$~\cite{Taub}:
\begin{equation}
{\cal L}_u=-\sqrt{-g}\Lambda(u^\mu u_\mu-1).
\label{vel}
\end{equation}
However, unlike in~\cite{Taub}, we do not treat the four-velocity as a variable with respect to which the total action is varied.
The dynamical variables are: the metric tensor, the affine connection, and the Lagrange multiplier $\Lambda$.

We also assume that the four-velocity field couples to a geometrical vector.
First, we consider that $u_\mu$ couples to the trace of the nonmetricity tensor $N^\rho_{\phantom{\rho}\rho\mu}$:
\begin{equation}
{\cal L}_N=\sqrt{-g}kN^\rho_{\phantom{\rho}\rho\mu}u^\mu,
\label{coup1}
\end{equation}
where $k$ is a coupling strength.
If we associate $k$ with $-\frac{i\alpha}{2}\rho_e$, where $\rho_e$ is the electric charge density, Eq.~(\ref{coup1}), because of Eq.~(\ref{prop1}), reproduces the electromagnetic coupling $\sim\rho_e A_\mu u^\mu$.
For the total matter Lagrangian ${\cal L}_m={\cal L}_u+{\cal L}_N$, Eq.~(\ref{Pi}) gives the hypermomentum density
\begin{equation}
\Pi_{\phantom{\mu}\rho\phantom{\nu}}^{\mu\phantom{\rho}\nu}=4\kappa\sqrt{-g}k\delta^\mu_\rho u^\nu,
\label{den1}
\end{equation}
which, using Eqs.~(\ref{Max2}) and~(\ref{prop3}), agrees with $j^\nu=\rho_e u^\nu$.
Substituting Eq.~(\ref{den1}) to~(\ref{sol}) gives $V^\rho_{\phantom{\rho}\mu\nu}=0$.
Therefore, using Eqs.~(\ref{LagGrav}) and~(\ref{Ein2}), we obtain the general-relativistic ${\cal L}_g=-\frac{R_g}{2\kappa}\sqrt{-g}$.
The torsion vector~(\ref{tor}) is proportional to the electromagnetic potential: $S_\nu=-\frac{3i}{8\alpha}A_\nu$~\cite{Ham}.

Varying the total action corresponding to the sum of:~(\ref{LagGrav}), (\ref{LagEM}), (\ref{vel}), and~(\ref{coup1}), with respect to $g^{\mu\nu}$ and $\Lambda$, and then using $u^\mu u_\mu=1$, give
\begin{eqnarray}
& & \kappa^{-1}G_{\mu\nu}=-\alpha^2\Bigl(\frac{1}{4}Q_{\rho\sigma}Q^{\rho\sigma}g_{\mu\nu}-Q_{\mu\rho}Q_\nu^{\phantom{\nu}\rho}\Bigr) \nonumber \\
& & +2\Lambda u_\mu u_\nu-A_{(\mu}j_{\nu)}+\frac{1}{2}g_{\mu\nu}A_\rho j^\rho,
\label{Ein3}
\end{eqnarray}
where $G_{\mu\nu}=R_{\mu\nu}^{(g)}-\frac{1}{2}R_g g_{\mu\nu}$ is the Einstein tensor.
Therefore the Lagrange multiplier $\Lambda$ represents the energy density $\rho_m$ for matter in the form of a pressureless perfect fluid:
\begin{equation}
\Lambda=\frac{\rho_m}{2}.
\label{dens}
\end{equation}

The Bianchi identity: $G^{\mu\nu}_{\phantom{\mu\nu}:\nu}=0$, yields the covariant conservation of the total energy-momentum tensor given by the right-hand side of Eq.~(\ref{Ein3}).
Neglecting the terms with $A_\mu j_\nu$ (which corresponds to regarding a fluid as an ensemble of non-interacting particles~\cite{LL}), and using Eqs.~(\ref{as}), (\ref{Max1}), (\ref{prop1}), (\ref{prop3}), and~(\ref{dens}), we obtain $\rho_m u^\nu u^\mu_{\phantom{\mu}:\nu}=\rho_e F^{\mu\nu}u_\nu$, or, since $\frac{\rho_e}{\rho_m}=\frac{q}{m}$, the Lorentz equation of motion for a test particle of mass $m$ and charge $q$: $mu^\nu u^\mu_{\phantom{\mu}:\nu}=qF^{\mu\nu}u_\nu$.

We now consider that $u_\mu$ couples to the the torsion vector:
\begin{equation}
{\cal L}_S=\sqrt{-g}k S_\mu u^\mu.
\label{coup2}
\end{equation}
Eq.~(\ref{Pi}) for the total matter Lagrangian ${\cal L}_m={\cal L}_u+{\cal L}_S$ gives the hypermomentum density
\begin{equation}
\Pi_{\phantom{\mu}\rho\phantom{\nu}}^{\mu\phantom{\rho}\nu}=2\kappa\sqrt{-g}k\delta^{[\mu}_\rho u^{\nu]},
\label{den2}
\end{equation}
and, using Eqs.~(\ref{Max2}) and~(\ref{prop3}), the current vector
\begin{equation}
j^\nu=\frac{3i}{8\alpha}ku^\nu.
\label{cur}
\end{equation}
Thus the coupling $k$ is now related to $\rho_e$ via $k=-\frac{8i\alpha}{3}\rho_e$.
Substituting Eq.~(\ref{den2}) to~(\ref{sol}) gives $V^\rho_{\phantom{\rho}\mu\nu}=\frac{\kappa k}{8}(3g_{\mu\nu}u^\rho-2u_{(\mu}\delta^\rho_{\nu)})$, which, after putting into Eqs.~(\ref{LagGrav}) and~(\ref{Ein2}), yields
\begin{equation}
{\cal L}_g=-\frac{R_g}{2\kappa}\sqrt{-g}+\frac{3\kappa k^2}{64}u^\mu u_\mu \sqrt{-g}.
\label{scal}
\end{equation}
The torsion vector~(\ref{tor}) becomes a linear combination of the electromagnetic potential and the electromagnetic current: $S_\nu=-\frac{3i}{8\alpha}A_\nu+\frac{i\kappa\alpha}{4}j_\nu$, turning the Lagrangian density~(\ref{coup2}) into
\begin{equation}
{\cal L}_S=-\sqrt{-g}\Bigl(A_\mu j^\mu+\frac{3\kappa k^2}{32}u^\mu u_\mu\Bigr).
\label{tors}
\end{equation}

Varying the total action corresponding to the sum of:~(\ref{LagEM}), (\ref{vel}), (\ref{scal}), and~(\ref{tors}), with respect to $g^{\mu\nu}$ and $\Lambda$, and then using $u^\mu u_\mu=1$, give
\begin{eqnarray}
& & \kappa^{-1}G_{\mu\nu}=-\alpha^2\Bigl(\frac{1}{4}Q_{\rho\sigma}Q^{\rho\sigma}g_{\mu\nu}-Q_{\mu\rho}Q_\nu^{\phantom{\nu}\rho}\Bigr)+2\Lambda u_\mu u_\nu \nonumber \\
& & -A_{(\mu}j_{\nu)}+\frac{1}{2}g_{\mu\nu}A_\rho j^\rho+\frac{3}{64}\kappa k^2(2u_\mu u_\nu+g_{\mu\nu}).
\label{Ein4}
\end{eqnarray}
This equation differs from Eq.~(\ref{Ein3}) by the last term that represents a perfect fluid with a negative energy density $\rho_n=-\frac{\kappa n^2}{24}$ (since $k$ is imaginary), where $n=\frac{\rho_e}{e}$ is on the order of the charged-particle concentration, and the equation of state $p=-\frac{\rho_n}{3}$.
Even for matter composing white dwarfs ($\rho_m\sim10^6 g/cm^3$) $\rho_n$ is negligible: $\rho_n/\rho_m\sim10^{-44}$.
Therefore the Bianchi identity leads again to the Lorentz equation.
If $n$ is at the Planck scale, the total energy density does not satisfy the strong energy condition, which may prevent the formation of classical spacetime singularities.

{\em Discussion}.
For a linear connection, not restricted to be metric compatible and symmetric, there are five possible modifications of the Maxwell equations: $g^{\nu\rho}F_{\nu\mu;\nu}=j_\mu$, $F^\nu_{\phantom{\nu}\mu;\nu}=j_\mu$, $F^{\nu\mu}_{\phantom{\nu\mu};\nu}=j^\mu$, $g^{\nu\rho}F_{\nu\phantom{\mu};\rho}^{\phantom{\nu}\mu}=j^\mu$, and the metrically modified $g^{\nu\rho}F_{\nu\mu:\nu}=j_\mu$~\cite{Coley}.
The experimentally confirmed conservation laws of electric charge and magnetic flux indicate that the last possibility, which also results from the differential-form and metric-free formulations of electrodynamics, is physical~\cite{Van}.
Regarding the tensor of homothetic curvature as the geometrical quantity representing the electromagnetic field tensor in the metric-affine gravity, and using the simplest form of the Lagrangian that depends of the tensor of homothetic curvature, automatically lead to the metrically modified Maxwell equations, supporting geometrization of electromagnetism.

In the presence of the gravitational field we correct the derivative by introducing the affine connection, while in the presence of the electromagnetic field we introduce the electromagnetic potential.
Therefore it seems natural to assume that the electromagnetic potential is related to the connection~\cite{PO} rather than the metric as in earlier unified field theories~\cite{Weyl,Kaluza,Schrod}.

The presented geometrical formulation of classical electromagnetism predicts two new phenomena that distinguish it from the Einstein-Maxwell theory.
First, if we assume that the four-velocity couples to the torsion vector, the energy density at extremely high particle concentrations becomes negative which may prevent the formation of singularities.
This possibility will be studied elsewhere.
Second, the coupling between spinors and the electromagnetic field represented by the tensor of homothetic curvature gives, as in theories where torsion couples to spin, the Heisenberg-Ivanenko equation instead of the Dirac equation~\cite{PO,st,NikoG}.

\end{document}